\documentclass[envcountsame]{llncs}
\usepackage{times}
\usepackage{graphicx} 
\usepackage{epsfig}
\usepackage{stmaryrd}
\usepackage{latexsym}
\usepackage{amssymb}
\usepackage{daytime}
\usepackage{eepic}
\usepackage{url}
\usepackage{verbatim}
\usepackage{cite}

\def\thisPaperTitle{On Spatial Conjunction as Second-Order Logic}

\usepackage{defs}
\title{\thisPaperTitle}

\author{Viktor Kuncak and Martin Rinard}
\institute{{\tt $\{$vkuncak,rinard$\}$@csail.mit.edu}\\
MIT CSAIL Technical Report 970, \ October 2004 \\
Computer Science and AI Lab \\
Massachusetts Institute of Technology \\
Cambridge, MA 02139, USA
}

\begin{document}

\sloppy

\maketitle

\begin{abstract}
Spatial conjunction is a powerful construct for reasoning
about dynamically allocated data structures, as well as
concurrent, distributed and mobile computation.  While
researchers have identified many uses of spatial
conjunction, its precise expressive power compared to
traditional logical constructs was not previously known.

In this paper we establish the expressive power of spatial
conjunction.  We construct an embedding from first-order
logic with spatial conjunction into second-order logic, and
more surprisingly, an embedding from full second order logic
into first-order logic with spatial conjunction.  These
embeddings show that the satisfiability of formulas in
first-order logic with spatial conjunction is equivalent to
the satisfiability of formulas in second-order logic.

These results explain the great expressive power of spatial
conjunction and can be used to show that adding unrestricted
spatial conjunction to a decidable logic leads to an
undecidable logic.  As one example, we show that adding
unrestricted spatial conjunction to two-variable logic leads
to undecidability.

On the side of decidability, the embedding into second-order
logic immediately implies the decidability of first-order
logic with a form of spatial conjunction over trees.  The
embedding into spatial conjunction also has useful
consequences: because a restricted form of spatial
conjunction in two-variable logic preserves decidability, we
obtain that a correspondingly restricted form of
second-order quantification in two-variable logic is
decidable.  The resulting language generalizes the
first-order theory of boolean algebra over sets and is
useful in reasoning about the contents of data structures in
object-oriented languages.

\end{abstract}
\paragraph{Keywords:} program specification, separation logic, spatial conjunction, second-order logic, shape analysis, two-variable logic





\section{Introduction}

Separation logic with spatial conjunction operator was
introduced as a technique for local reasoning about shared
mutable data
structures~\cite{Reynolds00IntuitionisticReasoningMutable,
IshtiaqOHearn01BIAssertionLanguage} and proved to be
remarkably effective~\cite{Reynolds02SeparationLogic,
BirkedalETAL04LocalReasoningCopyingCollector,
OHearnETAL01LocalReasoning,
CalcagnoETAL02DecidingValiditySpatialLogicTrees,
CalcagnoETAL01ComputabilityComplexityResultsSpatialAssertion,
BerdineETAL04DecidableFragmentSeparationLogic}.  Similar
constructs are present in formalisms based on process
calculi and ambient calculi
~\cite{CairesCardelli03SpatialLogicConcurrencyPartI,
Caires04BehavioralSpatialObservationsLogicPiCalculus,
LozesCaires04EliminationQuantifiersUndecidabilitySpatialLogicsConcurrency,
CardelliGordon00AnytimeAnywhere, Ghelli02SpatialLogicMFPS,
CardelliGhelli01QueryLanguageAmbientLogic,
LucaGhelli02SpatialLogicQueryingGraphs}.

Despite the increasing range of results and applications of
separation logic, the precise expressive power of spatial
conjunction constructs is often not known.  For example, the
authors in~\cite{Ghelli02SpatialLogicMFPS,
DawarETAL04ExpressivenessComplexityGraphLogic} use the
formalism of edge-labelled multigraphs and observe great
expressive power of spatial logic for describing paths in a
graph, but suggest that the relationship with second-order
logic in this setting is not straightforward.

In~\cite{KuncakRinard04OnRecordsSpatialConjunctionRoleLogic,
KuncakRinard04GeneralizedRecordsRoleLogic} we
defined the notion of spatial conjunction for arbitrary
relational structures.  Our notion of spatial conjunction
splits relations into disjoint subsets and has a natural
semantics that works for relations of any arity.  The
interpretation over relational structures is an important
step in enabling the combination of spatial conjunction with
the traditional first-order and second-order
logics~\cite{Mendelson97IntroductionMathematicalLogic,
BarwiseFeferman85ModelTheoreticLogics, Hodges93ModelTheory}
and their fragments.  One such decidable fragment of
first-order logic that is useful for reasoning about the
heap is two-variable logic with counting~\cite{GraedelETAL97TwoVariableLogicCountingDecidable},
whose variable-free counterpart is role
logic~\cite{KuncakRinard03OnRoleLogic}.
In~\cite{KuncakRinard04OnRecordsSpatialConjunctionRoleLogic,
KuncakRinard04GeneralizedRecordsRoleLogic} we
present a combination of two-variable logic with spatial
conjunction defined on relational structures and show that
it is useful for specifying generalized records that
formalize role constraints~\cite{KuncakETAL02RoleAnalysis}.
To preserve the decidability of the notation,
\cite{KuncakRinard04GeneralizedRecordsRoleLogic} imposes the
following restriction on spatial conjunction: spatial
conjunction may only be applied to formulas of (counting)
quantifier nesting at most one.  Under this assumption, we
show that spatial conjunction can be eliminated using
syntactic operations on formulas, which means that spatial
conjunction not only preserves decidability, but leaves the
expressive power of two-variable logic with counting
unchanged.

Given the results
in~\cite{KuncakRinard04GeneralizedRecordsRoleLogic}, a
natural question to ask is: are we imposing an unnecessarily
strong restriction by not allowing application of spatial
conjunction to formulas with nested quantifiers; in
particular, what is the decidability of logic that allows
spatial conjunction of formulas with two nested quantifiers?
The present paper gives an answer to this question: we
establish that allowing spatial conjunction for formulas
with nested quantifiers leads to an undecidable logic.  This
undecidability result turns out to be a consequence of an
unexpectedly fundamental connection: \emph{spatial
conjunction can represent second-order quantification}.  We
obtain a striking contrast on the expressive power of logic
depending on the use of spatial conjunction: if applied to
formulas with no nesting of first-order counting
quantifiers, the result is still two-variable logic with
counting; if applied to formulas with nested first-order
quantifiers, the resulting formulas can represent
second-order formulas.  This contrast can be viewed as a
justification for the restriction imposed in
\cite{KuncakRinard04GeneralizedRecordsRoleLogic}.

Because it applies to both decidable and undecidable logics,
the embedding of second-order logic into spatial conjunction
yields not only undecidability, but also decidability
results.  Using the restriction on the use of spatial
conjunction with the translation of second-order quantifiers
yields a decidable notation with second-order quantifiers.
This notation leads to a generalization of boolean algebra
of sets to two-variable logic with counting extended with a form of
second-order quantification; such notation is useful for reasoning
about data structure
abstractions~\cite{LamETAL04OurExperienceModularPluggableAnalyses,
LamETAL04HobProjectWebPage}.

We also note that graph reachability, inductive definitions,
spatial implication, and a parameterized version of spatial
conjunction are all expressible in second-order logic.  An
interesting consequence of the embedding of second-order
quantifiers into spatial conjunction is that all these
constructs are expressible using spatial conjunction alone.

Moreover, the converse embedding holds as well: spatial
conjunction is expressible in second-order logic.  Together,
these two results lead to a particularly simple
characterization: \emph{spatial conjunction and second-order
logic are equivalent} (see
Proposition~\ref{prop:spatialToSOL} and
Proposition~\ref{prop:SOLToSpatial} for the precise
formulation of this equivalence).

The translation from spatial logic to second-order logic
also has useful consequences.  Namely, if we restrict the
set of models to unions of trees, then monadic second-order
logic is decidable.  By translating restricted spatial logic
formulas to monadic second-order logic, we obtain that
spatial logic is decidable over trees as well.

In general, the equivalence for satisfiability between
spatial conjunction and second-order logic improves our
understanding of spatial conjunction and suggests that the
definition of spatial conjunction on relational structures
is a natural one.  While it is less surprising that
second-order logic can express the definition of spatial
conjunction (we have observed this already in the technical
report~\cite{KuncakRinard04OnRecordsSpatialConjunctionRoleLogic}),
we found it quite surprising that spatial conjunction in
first-order logic can express the entire second-order logic.
The idea of both directions of our translation is remarkably
simple, and this simplicity is reflected in the linear time
complexity of formula translations: translation of spatial
conjunction connectives into second-order logic mimics the
semantics of spatial conjunction in terms of the existence
of disjoint relations, and the translation from second-order
logic into spatial conjunction takes the advantage of the
non-determinism in splitting of the heap to simulate the
existential quantifier.

\paragraph{Contributions.}  We summarize the contributions of this paper as
follows.
\begin{enumerate}
\item We construct an equivalence-preserving translation of spatial 
conjunction into second-order quantifiers.  We then show
that this translation implies decidability of the
first-order logic with a spatial conjunction interpreted over
tree structures, when spatial conjunction splits only unary predicates.
\item We construct a satisfiability-preserving translation of 
second-order quantifiers into spatial conjunction, and
derive the following consequences:
\begin{enumerate}
\item first-order logic with spatial conjunction has the expressive
power of second-order logic, even if restricted to two
first-order variables, and even if spatial conjunction is
applied only to formulas of first-order quantifier nesting
at most two (similar result holds for parameterized spatial
conjunction that splits only \emph{unary} predicates: the resulting
logic is equivalent to \emph{monadic} second-order logic);
\item two-variable logic with counting extended with 
second-order quantifiers that apply only to formulas with
quantifier nesting at most one can be translated into
two-variable logic with counting, and is therefore decidable;
\item graph reachability, inductive definitions, spatial 
implication, and generalized spatial conjunction are all
expressible using first-order logic with spatial
conjunction.
\end{enumerate}
\end{enumerate}


\section{Preliminaries}

In this section we present our definitions of relational
structures as well as the semantics of second-order logic and
spatial conjunction.

\subsection{Relational Structures}
\label{sec:relStruct}

\begin{figure}
\begin{equation*}
\begin{array}{rcl}
  \Var & - & \mbox{fixed set of first-order variables} \mnls
  \Sigma & - & \mbox{finite vocabulary of 
                     relational symbols (2nd order variables)} \mnls
  \Sigmaof{k} & = & \{ \PS{k}_1, \PS{k}_2, \ldots, \PS{k}_{C_k} \} \\
  & & 
        \mbox{relational symbols of arity $k$}, \ar{\PS{k}}=k \mnls
  \Sigma &=& \cup_{k=0}^C \Sigmaof{k},\ C - \mbox{maximal arity} \mnls
  U & - & \mbox{fixed universe (domain) of relational structures} \\
  e & - & \mbox{relational structure (interpretation)} \\
  e & : & (\Var \to U) \cup \bigcup_k (\Sigmaof{k} \to 2^{U^k}) \\
  \ \\
  \mtr{x_1=x_2}e &\eqdef& (e(x_1)=e(x_2)) \mnls
  \mtr{\PS{k}(x_1,\ldots,x_k)}e &\eqdef& (e(x_1),\ldots,e(x_k)) \in e(\PS{k}) \mnls
  \mtr{F_1 \land F_2}e &\eqdef& \mtr{F_1}e \land \mtr{F_2}e \mnls
  \mtr{\lnot F}e &\eqdef& \lnot \mtr{F}e \mnls
  \mtr{\exists x.F}e &\eqdef& \exists v.\ \mtr{F}(e[x:=v]) \mnls
  \mtr{\exists \PS{k}.F}e &\eqdef& \exists r \subseteq U^k.\ \mtr{F}(e[\PS{k}:=r])
\end{array}       
\end{equation*}
\caption{Semantics (Interpretation) of Second-Order Formulas in Relational Structures
\label{fig:secondInterp}}
\end{figure}

Figure~\ref{fig:secondInterp} presents the semantics of
second-order logic formulas in relational structures, which
is mostly standard.  We use $\Var$ to denote first-order
variables with typical representatives $x$, $x_i$.  We use
$\Sigma$ to denote second-order variables (predicates), with
a typical representative $P$, or $\PS{k}$ when we wish to
specify that the predicate symbol has arity $k$;
alternatively we write $\ar{P}=k$.

For convenience we fix a universe $U$ of all relational
structures in a given context; we assume $U$ is countable,
but the cardinality of $U$ does not play an important role
for us.  A relational structure, denoted $e$, is a valuation
for first-order and second-order variables.  As in
first-order logic, for a first-order variable $x$, $e(x) \in
U$ is an element of the domain, and for a predicate symbol
$P$ of arity $\ar{P}=k$, $e(P) \subseteq U^{\ar{k}}$ is a
relation of arity $k$.  In this way we merge
the model and the variable assignment, which makes it natural to
define second-order quantification as in
Figure~\ref{fig:secondInterp}.  If $v$ is a first-order or
second-order variable, we use the standard notation
$e[v:=a]$ to denote the updated relational structure such
that $e[v:=a](v) = a$ and $e[v:=a](v_1) = e(v_1)$ for $v_1
\neq v$.  We treat equality in formulas as a logical symbol and
interpret it in the standard way.

\subsection{Spatial Conjunction}
\label{sec:spatialDef}

\begin{figure*}
  \begin{equation*}
    \begin{array}{rcl}
      \mtr{F_1 \spand F_2}e &\eqdef& \exists e_1,e_2.\
        \splitStruct{\Sigma}{e}{e_1,e_2}\ \land \
        \mtr{F_1}e_1 \land \mtr{F_2}e_2 \mnl
      \splitStruct{\sigma}{e}{e_1,e_2} &\eqdef&
      \bigwedge_{r \in \sigma} \splitRel{e(r)}{e_1(r),e_2(r)} \ \land \mnls
      & & 
      \bigwedge_{r \in \Sigma \setminus \sigma}
        (e_1(r) = e(r) \land e_2(r) = e(r))  \mnl
      \splitRel{r}{r_1,r_2} &\eqdef&
        (r = r_1 \cup r_2 \ \land\ r_1 \cap r_2 = \emptyset)
    \end{array}
  \end{equation*}
  \centerline{\includegraphics[scale=0.4]{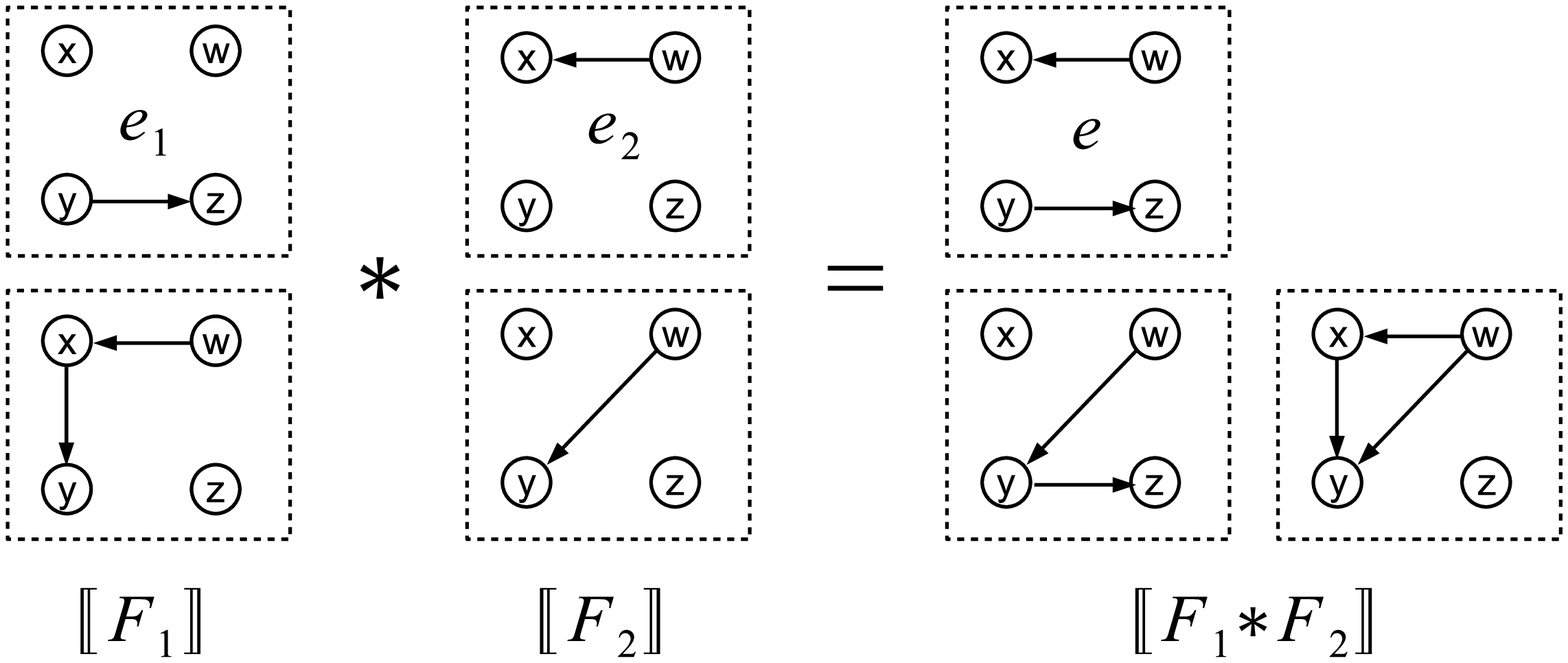}}
  \caption{Semantics of Spatial Conjunction $\spand$.\label{fig:spatialDef}}
 \begin{equation*}
   \begin{array}{rcl}
     \mtr{F_1 \spandof{\sigma} F_2}e & \eqdef & \exists e_1,e_2.\
       \splitStruct{\sigma}{e}{e_1,e_2}\ \land \
       \mtr{F_1}e_1 \land \mtr{F_2}e_2 \mnl
     F_1 \spand F_2 & \iff & F_1 \spandof{\Sigma} F_2
   \end{array}
 \end{equation*}
  \caption{Parameterized Spatial Conjunction $\spandof{\sigma}$
  \label{fig:genSpatialConjunction}}
\end{figure*}

Figure~\ref{fig:spatialDef} introduces our notion of spatial
conjunction, denoted $\spand$.  We illustrate the intuition
behind the definition of $\spand$ in terms of combining the
structures for which the formula is true.  Suppose $\Sigma =
\{ \PS{2} \}$ has only one binary relation symbol, so the
relational structures are graphs.  If $e_1$ is a structure
such that $\mtr{F_1}e_1$ and $e_2$ is a structure such that
$\mtr{F_2}e_2$, then if the edges of $e_1(\PS{2})$ and
$e_2(\PS{2})$ are disjoint, the structure with relation
$e(\PS{2}) = e_1(\PS{2}) \cup e_2(\PS{2})$ satisfies
$\mtr{F_1 \spand F_2}e$.  In general, there is one relation
$e$ for each pair of models $e_1$ and $e_2$ that can be
combined.  There are three models of $F_1 \spand F_2$ in
Figure~\ref{fig:spatialDef}; there is only one pair of
relations that cannot be combined, because of an overlapping
edge from $w$ to $x$.

The definition of spatial conjunction in
Figure~\ref{fig:spatialDef} is identical to the one we use
in~\cite{KuncakRinard04GeneralizedRecordsRoleLogic,
KuncakRinard04OnRecordsSpatialConjunctionRoleLogic}.  In our
setup, similarly to other notions of spatial
conjunction~\cite{LucaGhelli02SpatialLogicQueryingGraphs,
IshtiaqOHearn01BIAssertionLanguage}, a formula $F_1 \spand
F_2$ holds for a relational structure if and only if the
structure can be split into two disjoint structures where
$F_1$ holds for the first component and $F_2$ holds for the
second component.  The difference
with~\cite{LucaGhelli02SpatialLogicQueryingGraphs} is that
we use general relational structures which correspond to
labelled graphs as opposed to multigraphs.  Our notion of
splitting of relational structures, given by condition
$\splitStruct{\sigma}{e}{e_1,e_2}$, reduces to partitioning
each relation in $\sigma$.  For the definition of spatial
conjunction $\spand$ we let $\sigma = \Sigma$ where $\Sigma$
is the set of all relation symbols; it is also natural to
allow a generalized spatial conjunction $\spandof{\sigma}$
in Figure~\ref{fig:genSpatialConjunction} that takes the set
of predicate symbols $\sigma$ as an argument, then splits
relations in $\sigma$ and preserves the relations in $\Sigma
\setminus \sigma$.  For example, if we let $\sigma =
\Sigmaof{1}$, then the conjunction $\spandof{\sigma}$ splits
only unary relations.  The results of this paper imply that
$\spandof{\Sigma}$ corresponds to full second-order logic,
whereas $\spandof{\Sigmaof{1}}$ corresponds to monadic
second-order logic.

Our definition of spatial conjunction above is
not the only one possible, but there are several reasons to
consider it as a natural definition of spatial conjunction:
\begin{itemize}
\item Our definition is close to the definition of
\cite{IshtiaqOHearn01BIAssertionLanguage}.  A relational
structure can represent a store by modelling each store
location as a pair of an object and one of the finitely
many predicate symbols; this view is appropriate for type-safe
languages such as Java, ML, and O'Caml.
\item The only difference compared
to~\cite{LucaGhelli02SpatialLogicQueryingGraphs} is that we
use relations as sets of tuples
where~\cite{LucaGhelli02SpatialLogicQueryingGraphs} uses
multigraphs as multisets of tuples; we believe that our
results can provide useful insight into languages such
as \cite{LucaGhelli02SpatialLogicQueryingGraphs} as well.
\item With the appropriate definition of spatial implication
$\spimpl$ (Figure~\ref{fig:spatialImpl}) corresponding to
conjunction $\spand$, our model validates the axioms of
bunched
implications~\cite{OHearnPym99LogicBunchedImplications,
IshtiaqOHearn01BIAssertionLanguage}.
\item We can naturally describe concatenation of generalized 
records~\cite{KuncakRinard04OnRecordsSpatialConjunctionRoleLogic,
KuncakRinard04GeneralizedRecordsRoleLogic}, which cannot be
expressed using standard logical operations.
\end{itemize}

The main claim of this paper is that our notion of spatial
conjunction is equivalent for satisfiability to second-order
quantification.  This equivalence can be viewed as another
argument in favor of the definitions we adopt.  We proceed
to demonstrate both directions of the equivalence, and then
present some consequences of the result.



\section{Representing Spatial Conjunction $\spand$ in Second-Order Logic}
\label{sec:secondSpatial}

In this section we give a translation from the first-order
logic with spatial conjunction to second-order logic.  The
consequence of this translation is an upper bound on the
expressive power of spatial conjunction.  Because our
translation applies to all relational structures, if we
restrict the set of relational structures so that
second-order logic becomes decidable, then the corresponding
spatial logic is decidable on the restricted set of
structures as well.

\begin{figure}
\begin{equation*}
  \begin{array}{l}
    \begin{array}{rcl}
      \sigma &=& \{ P_1,\ldots, P_n \} \mnl
  \multicolumn{3}{l}{\mbox{spatial conjunction elimination:}} \\
      \trps{F' \spandof{\sigma} F''} &=& 
        \exists P'_1,\ldots,P'_n,P''_1,\ldots,P''_n. \mnls
        & &
        \bigwedge_{i=1}^n \synSplitRel{P_i}{P'_i,P''_i} \ \land \mnls
        & &
        F'[P_i:=P_i']_{i=1}^n \ \land\
        F''[P_i:=P_i'']_{i=1}^n
        \mnl
      \synSplitRel{P}{P',P''} &=& \forall x_1,\ldots,x_k.\\
        &&
        \qquad
        \begin{array}[t]{l}
           (P(x_1,\ldots,x_k) \miff P'(x_1,\ldots,x_k) \lor P''(x_1,\ldots,x_k)) \ 
            \land \\
           \lnot ( P'(x_1,\ldots,x_k) \land P''(x_1,\ldots,x_k)) 
        \end{array} \\
       && k = \ar{P} = \ar{P'} = \ar{P''} \mnl
    \end{array} \\
    \begin{array}{rcl}
  \multicolumn{3}{l}{\mbox{recursive translation function:}} \\
  \rtrps{x_1=x_2} & = & (x_1=x_2) \mnls
  \rtrps{\PS{k}(x_1,\ldots,x_k)} & = & \PS{k}(x_1,\ldots,x_k) \mnls
  \rtrps{F_1 \land F_2} & = & \rtrps{F_1} \land \rtrps{F_2} \mnls
  \rtrps{\lnot F} & = & \lnot \rtrps{F} \mnls
  \rtrps{\exists x.F} & = & \exists x. \rtrps{F} \mnls
  \rtrps{\exists \PS{k}.F} & = & \exists \PS{k}.\ \rtrps{F} \mnls
  \rtrps{F_1 \spand F_2} & = &
     \trps{\rtrps{F_1} \spandof{\Sigma} \rtrps{F_2}} \mnl
  \rtrps{F_1 \spandof{\Sigmaof{1}} F_2} & = &
     \trps{\rtrps{F_1} \spandof{\Sigmaof{1}} \rtrps{F_2}} \mnl
   \end{array} \\
     \begin{array}{rcl}
  \multicolumn{3}{l}{\mbox{translation correctness lemmas:}} \\
   \mtr{F'[P_i := P'_i]_{i=1}^n} (e[P'_i := r_i]_{i=1}^n) & = &
   \mtr{F'} (e[P_i := r_i]_{i=1}^n) \qquad \mbox{for $P_i$ fresh in $F'$} \mnls
   \mtr{\trps{F}}e & = & \mtr{F}e \mnls
   \mtr{\rtrps{F}}e & = & \mtr{F}e
      \end{array}
  \end{array}
\end{equation*}
  \caption{Translation of Spatial Conjunction into 
           Second-Order Logic\label{fig:spatialToSOL}}
\end{figure}

\noindent
Figure~\ref{fig:spatialToSOL} presents the translation from
first-order logic extended with spatial conjunction into
second-order logic.  The translation directly mimics the
semantics of $\spand$ and follows from the fact that
second-order logic can essentially quantify over its entire
domain and can express disjointness of relations.  Indeed,
the truth value of a formula depends only on finitely many
first and second-order variables, and second-order logic can
quantify over each of these variables, which amounts to
quantification over relational structures.

The translation in Figure~\ref{fig:spatialToSOL} introduces
two fresh predicate symbols $P'_i, P''_i$ for each predicate
symbol $P_i$ and asserts that $P'_i$ and $P''_i$ split
$P_i$.  The translation then replaces the predicates $P_i$
with the corresponding predicates $P'_i$ in the first
formula $F'$, and replaces the predicates $P_i$ with the
predicate $P''_i$ in the second formula $F''$.  The
correctness of the translation follows from the definitions,
using lemmas in Figure~\ref{fig:spatialToSOL} and structural
induction.  We conclude the following.

\begin{proposition} \label{prop:spatialToSOL}
  If $F$ is a second-order logic formula potentially
  containing spatial conjunction, then $\rtrps{F}$ is an
  equivalent second-order logic formula without spatial
  conjunction; we have $\mtr{\rtrps{F}}e = \mtr{F}e$ for all
  relational structures $e$ that interpret $F$.  Moreover,
  if $F$ is a monadic second-order logic formula with
  $\spandof{\Sigmaof{1}}$ as the only spatial conjunction
  operator, then the resulting formula is a monadic
  second-order logic formula.
\end{proposition}


\section{Representing Second-Order Quantifiers using $\spand$}
\label{sec:spatialSecond}

This section shows that second-order quantifiers can be
represented using spatial conjunction.  Among the
consequences of this result are the fact that first-order
logic with spatial conjunction has the expressive power of
second-order logic (even if restricted to two first-order
variables where the spatial conjunction connects only
formulas of first-order quantifier nesting at most two),
that two-variable logic with counting extended with
second-order quantifiers that apply only to formulas with
quantifier nesting at most one is decidable, and that
inductive definitions, spatial implication, and generalized
spatial conjunction are expressible using first-order logic
with spatial conjunction.

\begin{figure}
  \begin{equation*}
    \begin{array}{l}
      \begin{array}{rcl}
        \bvtwo{F} & - & \mbox{second-order variables bound in } F \mnls
        \vtwo{F} & - & \mbox{second-order variables in } F \mnls
        F_0 & - & \mbox{a formula without spatial conjunction $\spand$} \mnls
        \mbox{assumption:} & &
        \parbox[t]{3in}{
          all bound variables in $F_0$ are mutually distinct and \\
          distinct from free variables in $F_0$} \\
        \ \\
        \allpreds & \equiv & \displaystyle 
            \forall x. \bigwedge_{Q \in \bvtwo{F_0}} Q(x) \mnl
        \nonebut{P} & \equiv & \displaystyle 
            \forall x. \bigwedge_{Q \in \vtwo{F_0} \setminus \{P\}} \lnot Q(x) \\
      \end{array} \\ 
      \ \\
      \mbox{translation of second-order quantifier:} \\
      \trsp{\exists P.\, F} \ = \ \nonebut{P} \spand F \mnls
      \begin{array}{rcl}
        \multicolumn{3}{l}{\mbox{recursive translation function:}} \\
        \rtrsp{x_1=x_2} & = & (x_1=x_2) \mnls
        \rtrsp{\PS{k}(x_1,\ldots,x_k)} & = & \PS{k}(x_1,\ldots,x_k) \mnls
        \rtrsp{F_1 \land F_2} & = & \rtrsp{F_1} \land \rtrsp{F_2} \mnls
        \rtrsp{\lnot F} & = & \lnot \rtrsp{F} \mnls
        \rtrsp{\exists x.F} & = & \exists x. \rtrsp{F} \mnls
        \rtrsp{\exists \PS{k}.F} & = & 
          \trsp{\exists \PS{k}.\ \rtrsp{F}} \mnls
      \end{array} \\
      \mbox{final translation of a formula:} \mnls
      \ftrsp{F_0} \ = \ \allpreds \ \land\ \rtrsp{F_0} \mnl
      \mbox{translation correctness lemmas:} \mnls
      \begin{array}{rcl}
        \mtr{\exists P.\, F}e & = &
          \mtr{\trsp{\exists P.\, F}} (e[P:=U^{\ar{P}}]) \mnl
        \mtr{F_0}e & = &
          \mtr{\ftrsp{F_0}} (e[P:=U^{\ar{P}}]_{P \in \bvtwo{F_0}}) \mnl
        \exists e. \mtr{F_0}e & \iff &
        \exists e. \mtr{\ftrsp{F_0}} e
      \end{array}
    \end{array}
  \end{equation*}
  \caption{Translation of Second-Order Quantifiers into Spatial Conjunction
  \label{fig:transSecondSpatial}}
\end{figure}

Figure~\ref{fig:transSecondSpatial} presents the translation
of second-order quantifiers into spatial conjunction.  As in
the case of the converse translation in
Section~\ref{sec:secondSpatial}, the intuition behind the
translation is to exploit the semantics of spatial
conjunction in Figure~\ref{fig:genSpatialConjunction}.  This
time, however, we use the more complex operation---splitting
of relational structures---to simulate an existential
quantifier over relations, which leads to apparent
difficulties.  At first sight it appears that heap splitting
fails to have the effect of an existential quantifier over
a relation predicate, for two reasons:
\begin{enumerate}
\item splitting relational structures splits existing
relations, which means that the interpretations of 
relations in the resulting structure are subsets of the
interpretation of relations in the enclosing structure;
\item splitting of relational structures splits all relations,
and not just the interpretation of one predicate.
\end{enumerate}
We solve both of these problems when translating a formula
$F_0$ with second-order quantifiers, as follows.  We first
rename all bound second-order variables (denoted
$\bvtwo{F_0}$) to ensure that they are all distinct and that
they differ from the free variables in $F_0$.  In the
translated formula, even the bound second-order variables
$\bvtwo{F_0}$ become free second-order variables, which are
allowed in first-order logic.  To solve the first problem,
instead of considering all possible relational structures
$e$, we consider only those relational structures that map
the variables $\bvtwo{F_0}$ to full relations; we use the
conjunct $\allpreds$ to ensure that only such
structures are considered for the interpretation of the
final translated formula $\ftrsp{F_0}$.  We translate the
formula using the recursive translation function denoted
$\trsp{F}$, which walks the formula tree and applies the
translation of the existential quantifier.  The translation
of the existential quantifier, denoted $\trsp{\_}$, replaces
the quantifier $\exists P.\, F$ with the formula
$\nonebut{P} \spand F$.  The spatial conjunct $\nonebut{P}$
solves the second problem above, by asserting that all
relations other than $P$ are empty, and leaving the
interpretation of relation $P$ unconstrained.  As a result,
the interpretation of $P$ in $F$ is arbitrary, achieving the
effect of existential quantification, and the
interpretations of the remaining quantifiers remain the
same, as desired.

Soundness of the translation in
Figure~\ref{fig:transSecondSpatial} is given by
equisatisfiability, or equivalence on a reasonably
restricted class of structures, as summarized by the
following proposition.
\begin{proposition} \label{prop:SOLToSpatial}
  Let $F_0$ be a second-order logic formula in which each
  bound variable is distinct from all other variables in
  $F_0$.  Then $\ftrsp{F_0}$ is a formula in first-order
  logic with spatial conjunction, such that $F_0$ has a
  model if and only if $\ftrsp{F_0}$ has a model.  Moreover,
  if $e$ ranges over structures that assign full relations
  to predicate symbols not free in $F_0$, then the
  transformation is equivalence preserving, that is,
  $\mtr{F_0}e$ if and only if $\mtr{\ftrsp{F_0}}e$.
  Finally, if all second-order quantifiers in $F_0$
  are monadic, then $F_0$ can be translated into formula
  containing only $\spandof{\Sigmaof{1}}$ instead of
  $\spand$.
\end{proposition}

\section{Consequences of the Equivalence}

This section presents the consequences of the equivalence
between spatial conjunction and second-order quantification.

\subsection{Spatial Conjunction on Tree Structures is Decidable}

This section summarizes one interesting consequence of the
equivalence between spatial conjunction and second-order
logic with respect to tree structures.

Let us restrict our attention to relational structures that
interpret predicates of arity at most two.  Such relational
structures correspond to graphs with labelled nodes and
edges.  We say that a relational structure is a
\emph{forest} if the directed graph obtained by erasing all
labels is a directed forest, where by a directed forest we
mean a directed graph with no cycles where each node has an
in-degree at most one.  We then have the following
lemma.
\begin{lemma}
  If $e$ is a forest, and $\splitStruct{\Sigma}{e}{e_1,e_2}$
  holds, then both $e_1$ and $e_2$ are forests.
\end{lemma}
The previous lemma easily follows by contraposition: if
$e_1$ or $e_2$ have a cycle so does $e$, and if $e_1$ or
$e_2$ have a node with in-degree two or more, so does $e$.
This lemma implies that, when evaluating the meaning
$\mtr{F}e$ of formula in first-order logic with spatial
conjunction, it suffices to restrict the top-level structure
$e$ to be a forest for all structures occurring in the
semantics of subformulas of $F$ to be forests, which means
that the semantics of spatial conjunction over forests is
equivalent to the semantics in
Figure~\ref{fig:secondInterp}.  Using
Proposition~\ref{prop:spatialToSOL} we then obtain as a
special case $\mtr{F}e \iff \mtr{\rtrps{F}}e$.  By
decidability of monadic second-order logic over
trees~\cite{ComonETAL97Tata}, we conclude the following.
\begin{proposition}
  The satisfiability (and therefore the validity) problem of
  first-order logic extended with spatial conjunction 
  $\spandof{\Sigmaof{1}}$ is decidable.
\end{proposition}


\subsection{Undecidable Extension of Two-Variable Logic}
\label{sec:undecExt}

This section notes a consequence of
Proposition~\ref{prop:SOLToSpatial} on extensions of
decidable fragments of first-order logic with spatial
conjunction.  It is motivated by the following fact, proven
in \cite{KuncakRinard04GeneralizedRecordsRoleLogic}:
\begin{fact} \label{fact:previousCTwo}
  Two variable logic with counting extended with spatial conjunction
  on formulas with no nested counting quantifiers is decidable.
\end{fact}
A natural question to ask is: what is the decidability of
the notation if we allow spatial conjunction of formulas
with quantifier nesting two or more.  The answer is that the
resulting notation is undecidable.  Namely, if we have only
binary relation symbols, we obtain a logic equivalent to
full second-order logic, and already first-order logic in
the language with binary relation symbols is undecidable.

The reason for obtaining second-order logic when allowing
spatial conjunction of formulas with nested quantifiers is
that it is possible to simulate first-order quantifiers
using second-order quantifiers.  We can represent a
first-order variable such as $x$ by a second-order variable
$P_x$ bounded by the property $\exists_1 z. P_x(z)$, and
then replace each binary relation symbol $f(x,y)$ with a
formula of the form
\begin{equation} \label{eqn:correlation}
  \forall u.\forall v.\ P_x(u) \land P_y(v) \implies f(u,v),
\end{equation}
which uses only two first-order variables and has quantifier
nesting of two.  Similarly, the use of a unary relation
symbol $P(x)$ can be replaced by $\forall u.\, P_x(u) \implies
P(u)$.  

Now consider a second-order logic formula with binary and
unary relation symbols and no restrictions on the number of
first-order variables.  As described above, we can reduce
such formula to an equisatisfiable formula that uses only
two first-order variables.  We can then apply the
translation in Figure~\ref{fig:transSecondSpatial} to
eliminate second-order quantifiers.  Because formulas
$\allpreds$ and $\nonebut{P}$ have the quantifier depth at
most one, the result is a formula with spatial conjunction
that is applied to quantifiers of depth at most two and that
uses at most two first-order variable names.  Moreover, the
resulting formula is equisatisfiable by
Proposition~\ref{prop:SOLToSpatial}.  Because the
satisfiability of second-order logic formulas is
undecidable, the translation of second-order logic formulas
into formulas with spatial conjunction implies
undecidability of formulas with spatial conjunction applied
to formulas with quantifiers depth of two or more.
\begin{proposition} \label{prop:manyUndec}
  Two variable logic with counting extended with spatial
  conjunction $\spand$ on formulas with quantifier nesting at most
  two is undecidable.  The result applies to spatial
  conjunction $\spandof{\Sigmaof{1}}$ as well.
\end{proposition}


\subsection{Decidable Second-Order Quantification in Two-Variable Logic}

We next state a positive consequence of the
Fact~\ref{fact:previousCTwo} and
Proposition~\ref{prop:SOLToSpatial}.
\begin{proposition} \label{prop:secondDecid}
  Two variable logic with counting extended with
  second-order quantification on formulas with no nested
  counting first-order quantifiers is decidable.
\end{proposition}
Just like the previous Proposition~\ref{prop:manyUndec},
Proposition~\ref{prop:secondDecid} follows from applying the
translation in Figure~\ref{fig:transSecondSpatial} and
observing that the resulting formula has no nested
first-order quantifiers, and is equisatisfiable by
Proposition~\ref{prop:SOLToSpatial}.  Applying
Proposition~\ref{fact:previousCTwo}, we can decide the
satisfiability of the resulting formula, which gives the
satisfiability of the original formula as well.

To see why Proposition~\ref{prop:secondDecid} is
interesting, note that Proposition~\ref{prop:secondDecid}
places no restrictions on the number of second-order quantifiers used
on a formula with no nested first-order quantifiers.
Next, recall that monadic second-order logic of a set
(with no relation symbols) is just the first-order logic of
boolean algebra of sets, which is decidable by quantifier
elimination~\cite{Skolem19Untersuchungen} (for an overview
of quantifier elimination for boolean algebra see, for example,
\cite{KuncakRinard04FirstOrderTheorySetsCardinalityConstraintsDecidable}).
We therefore observe that the language permitted by
Proposition~\ref{prop:secondDecid} is a proper
generalization of boolean algebra of sets; it is a
generalization that allows stating set properties in a
neighborhood of a pair of objects given by two free
variables of a formula in two-variable logic with counting.

While we have found the first-order theory of boolean algebra of
sets to be useful for reasoning about the content of global 
data structures
\cite{LamETAL04OurExperienceModularPluggableAnalyses},
the generalization presented in this section allows
reasoning about sets that exist in the neighborhood of an
object denoted by a first-order variable.  In other words,
this specification language allows us to reason about the
content of data structures associated with individual
objects (which are common in object-oriented programming
languages), as opposed to just reasoning about global data
structures.

Comparing the results of this section and
Section~\ref{sec:undecExt}, we note the crucial role of the
restriction on quantifier nesting: with no nested
first-order quantifiers, it is not possible to use
second-order variables to simulate first-order variables
because it is not possible to establish the correlation of
the form~(\ref{eqn:correlation}).


\subsection{Expressing Inductive Definitions and Spatial Implication}

We next review the fact that inductive definitions (and
therefore transitive closure) are definable in second-order
logic.  This fact is of interest because it implies that
inductive definitions can be represented using spatial
conjunction, which leads to a surprising conclusion that
inductive definitions do not increase the expressive power
of first-order logic with spatial conjunction.  We similarly
observe that the spatial implication corresponding to spatial
conjunction is expressible in second-order logic and
therefore expressible using 
spatial conjunction.  All these consequences
follow from Proposition~\ref{prop:SOLToSpatial}.

\begin{figure}
  \begin{equation*}
    \begin{array}{l}
      \mtr{\letrec\ \PS{k}(x_1,\ldots,x_k) = F\ \iin\ G} \ \eqdef \
      \mtr{G[\PS{k} := \lfp{\PS{k}}{x_1,\ldots,x_k}{F}]} \mnl
      \mtr{\lfp{\PS{k}}{x_1,\ldots,x_k}{F}(y_1,\ldots,y_k)}e \ \eqdef \\
      \qquad
      (e(y_1),\ldots,e(y_k)) \in
      \lfpsem{\lambda r. \{ (v_1,\ldots,v_k) \mid 
              \mtr{F}e[\PS{k}:=r,x_1:=v_1,\ldots,x_k:=v_k]}
    \end{array}
  \end{equation*}
  \caption{Semantics of Inductive Definitions\label{fig:inductiveDef}}
\end{figure}

Figure~\ref{fig:inductiveDef} presents the semantics of
inductive definitions.  The
syntax of the least-fixpoint operator is
\begin{equation*}
  \lfp{\PS{k}}{x_1,\ldots,x_k}{F}(y_1,\ldots,y_k)
\end{equation*}
where $F$ is a formula that may contain new free variables
$\PS{k},x_1,\ldots,x_k$.  The meaning of the least-fixpoint
operator is that the relation which is the least fixpoint of
the monotonic transformation on predicates
\begin{equation*}
  (\lambda x_1,\ldots,x_k.\PS{k}(x_1,\ldots,x_k)) \mapsto (\lambda x_1,\ldots,x_k.F)
\end{equation*}
holds for $y_1,\ldots,y_k$.  To ensure the monotonicity
of the transformation on predicates, we require
that $\PS{k}$ occurs only positively in $F$.

\begin{figure}
  \begin{equation*}
    \begin{array}{l}
      \tris{\lfp{\PS{k}}{x_1,\ldots,x_n}{F}(y_1,\ldots,y_n)} = \\
      \quad \forall P.\, (\forall x_1,\ldots,x_n.\, 
               (F \miff P(x_1,\ldots,x_n))) \implies P(y_1,\ldots,y_n) \mnl
      \mbox{Soundness:} \mnl
      \mtr{\tris{\lfp{\PS{k}}{x_1,\ldots,x_n}{F}(y_1,\ldots,y_n)}}e =
      \mtr{\lfp{\PS{k}}{x_1,\ldots,x_n}{F}(y_1,\ldots,y_n)}e
    \end{array}
  \end{equation*}
  \caption{Expressing Least Fixpoint in Second-Order Logic\label{fig:lfpInSecond}}
\end{figure}

Figure~\ref{fig:lfpInSecond} shows that least fixpoint
operator is expressible in second-order logic.  The property
that $P$ is a fixpoint of $F$ is easily expressible.  To
encode that $y_1,\ldots,y_n$ hold for the {\em least}
fixpoint of $F$, we state that $y_1,\ldots,y_n$ hold for all
fixpoints of $F$, using universal second-order
quantification over $P$.

\begin{figure*}
 \begin{equation*}
   \begin{array}{rcl}
     \mtr{F' \spimpl F''}e & \eqdef & 
       \forall e',e''.\
       (\splitStruct{\Sigma}{e''}{e,e'}\ \land \
        \mtr{F'}e') \implies \mtr{F''}e'' \mnl
      \trps{F' \spimpl F''} &=& 
        \forall P'_1,\ldots,P'_n,P''_1,\ldots,P''_n. \mnls
        & &
        \big( 
        \bigwedge_{i=1}^n \synSplitRel{P''_i}{P_i,P'_i} \ \land \
        F'[P_i:=P_i']_{i=1}^n  \big) \implies \mnl
        & & \quad
        F''[P_i:=P_i'']_{i=1}^n
   \end{array}
 \end{equation*}
 \caption{Semantics of Spatial Implication\label{fig:spatialImpl}}
\end{figure*}

Figure~\ref{fig:spatialImpl} presents the semantics of the
spatial implication operation that along with spatial
conjunction $\spand$ validates the axioms of bunched
implications~\cite{OHearnPym99LogicBunchedImplications,
IshtiaqOHearn01BIAssertionLanguage}.
Figure~\ref{fig:spatialImpl} also presents the translation
of $\spimpl$ into second-order logic; the translation is
analogous to the translation of spatial conjunction in
Figure~\ref{fig:spatialToSOL}.  (As usual, the universal
quantifiers can be expressed using the existential
quantifier and negation.)

We summarize the results of this section as follows.
\begin{proposition}
  Graph reachability, inductive definitions, spatial
  implication, and generalized spatial conjunction are all
  expressible using first-order logic with spatial
  conjunction.
\end{proposition}



\section{Related Work}


The use of separation logic for reasoning about shared
mutable data structures started recently
\cite{Reynolds00IntuitionisticReasoningMutable,
IshtiaqOHearn01BIAssertionLanguage}
using ideas from
\cite{Burstall72SomeTechniquesProvingCorrectnessPrograms}
and proved very fruitful~\cite{Reynolds02SeparationLogic,
BirkedalETAL04LocalReasoningCopyingCollector,
OHearnETAL01LocalReasoning,
CalcagnoETAL02DecidingValiditySpatialLogicTrees,
CalcagnoETAL01ComputabilityComplexityResultsSpatialAssertion}.  Our notion
of spatial conjunction is
defined on relational structures rather than on mappings from
memory locations to values, but our model can represent a
location as a pair containing 1) an object and 2) one of the
finitely many field names.  Relational structures can
naturally represent memory models of languages with
destructive updates~\cite{SagivETAL02Parametric,
SagivETAL99Parametric, WilhelmETAL00ShapeAnalysis,
ManevichETAL04PartiallyDisjunctiveHeapAbstraction,
ManevichETAL02CompactlyRepresentingFirstOrderStructures,
LevAmiETAL00PuttingStaticAnalysisWorkVerification,
BoergerStaerk03AbstractStateMachines} and can also model
concurrency and temporal logic specifications
\cite{Yahav01VerifyingSafetyPropertiesConcurrentJava,
YahavETAL03VerifyingTemporalHeapPropertiesSpecifiedEvolutionLogic}.

Process
calculi~\cite{CairesCardelli03SpatialLogicConcurrencyPartI}
and ambient calculi~\cite{CardelliGordon00AnytimeAnywhere}
can reason about space and locality as well as concurrency;
these ideas also extend to graph-based
structures~\cite{Ghelli02SpatialLogicMFPS,
LucaGhelli02SpatialLogicQueryingGraphs}.  The results
closest to ours are
are~\cite{LucaGhelli02SpatialLogicQueryingGraphs,
Ghelli02SpatialLogicMFPS,
DawarETAL04ExpressivenessComplexityGraphLogic}; they are based
on edge-labelled multigraphs, and do not establish the
full equivalence with second-order logic.  Graph-based structures
in~\cite{LucaGhelli02SpatialLogicQueryingGraphs} are close
to the relational structures that we use, but use multisets
of edges instead of sets of edges.  Similarly to spatial
logic, type systems for reasoning about
aliasing~\cite{SmithETAL00AliasTypes,
WalkerMorrisett00AliasTypesRecursive,
FahndrichDeLine02AdoptionFocus,
DeLineFahndrich01EnforcingHighLevelProtocols} typically
contain join operators that combine independent portions of
store, although they are often based on linear types as
opposed to separation logic.

Our work clarifies the relationship between separation logic
and traditional first-order
logic~\cite{Mendelson97IntroductionMathematicalLogic} and
second-order
logic~\cite{BarwiseFeferman85ModelTheoreticLogics} and
explains surprising expressive power of spatial conjunction
without inductive definitions in expressing reachability
properties~\cite{Ghelli02SpatialLogicMFPS,
LucaGhelli02SpatialLogicQueryingGraphs}.
The understanding of separation logic in connection to other
formalisms is useful both for manual
reasoning~\cite{BirkedalETAL04LocalReasoningCopyingCollector}
and automated reasoning about programs with shared mutable
data
structures~\cite{CousotCousot79SystematicDesignProgramAnalysis,
Caires04BehavioralSpatialObservationsLogicPiCalculus,
SagivETAL99Parametric, SagivETAL02Parametric,
NielsonETAL01KleenesLogicEquality,
ManevichETAL04PartiallyDisjunctiveHeapAbstraction,
ManevichETAL02CompactlyRepresentingFirstOrderStructures,
LevAmiETAL00PuttingStaticAnalysisWorkVerification,
AikenETAL03CheckingInferringLocalNonAliasing,
Blanchet98EscapeAnalysis, Blanchet03EscapeAnalysis,
Moeller01PALE}.  
Decidability and complexity results of underlying logics and constraint
systems are particularly important for automated reasoning
\cite{CalcagnoETAL02DecidingValiditySpatialLogicTrees,
BenediktETAL99LogicForLinked,
BerdineETAL04DecidableFragmentSeparationLogic,
MelskiReps00InterconvertibilitySetConstraintsCFLReachability,
KodumalAiken04SetConstraintCFLReachabilityConnection,
CalcagnoETAL01ComputabilityComplexityResultsSpatialAssertion,
LozesCaires04EliminationQuantifiersUndecidabilitySpatialLogicsConcurrency}.

We have previously used the notion of spatial logic on
relational structures in
\cite{KuncakRinard04OnRecordsSpatialConjunctionRoleLogic,
KuncakRinard04GeneralizedRecordsRoleLogic} and presented
a novel use of spatial conjunction to describe
concatenation of generalized records.   In
\cite{KuncakRinard04OnRecordsSpatialConjunctionRoleLogic,
KuncakRinard04GeneralizedRecordsRoleLogic}
we take advantage
of the definition of spatial conjunction on relational structure to combine
it with a fragment of first-order logic: we present a decidable extension of
two-variable logic with counting and its variable-free version
role logic~\cite{KuncakRinard03OnRoleLogic}.  The encoding of
spatial conjunction in second-order logic appears in the technical report
\cite{KuncakRinard04OnRecordsSpatialConjunctionRoleLogic}; we have since
discovered the converse (and to us more surprising)
encoding.  The converse encoding gives justification to the
restriction
in~\cite{KuncakRinard04GeneralizedRecordsRoleLogic} by
showing that the absence of the restriction leads to an
undecidable, and in fact, extremely expressive, logic.
Moreover, the results of the present paper show how to use
second-order quantifiers in two-variable logic while
preserving decidability.  The resulting notation generalizes
the language of boolean algebra of sets, which we have found
useful in reasoning about data structure
abstractions~\cite{LamETAL04OurExperienceModularPluggableAnalyses,
LamETAL04HobProjectWebPage}.


\section{Conclusions}

In this paper we established the expressive power of spatial
conjunction by constructing an embedding from first-order
logic with spatial conjunction into second-order logic and
an embedding from full second order logic into first-order
logic with spatial conjunction.  These embeddings show that
the satisfiability of formulas in first-order logic with
spatial conjunction is equivalent to the satisfiability of
formulas in second-order logic.  This equivalence implies
new decidability and undecidability results for extensions
of two-variable logic with counting, decidability of
(unary-predicate) spatial logic over trees, and the fact
that inductive definitions, spatial implication, and a
parameterized spatial conjunction are all expressible using
first-order logic with spatial conjunction.  Finally, our
connection opens up the possibility of using second-order
logic as a unifying framework for integrating several
formalisms for reasoning about dynamic data structures:
spatial logic, monadic second-order logic on trees and
graphs, and three-valued structures.


\paragraph{Acknowledgements.} We thank the
participants of the Dagstuhl Seminar 03101 ``Reasoning about
Shape'' for useful discussions on separation logic, shape
analysis, and techniques for reasoning about mutable data
structures in general.  The consequences of the translation
from second-order logic to spatial conjunction for
two-variable logic with counting were crystallized in
discussion with Greta Yorsh.



\bibliographystyle{abbrv}
\bibliography{pnew}

\begin{thebibliography}{10}

\bibitem{AikenETAL03CheckingInferringLocalNonAliasing}
A.~Aiken, J.~S. Foster, J.~Kodumal, and T.~Terauchi.
\newblock Checking and inferring local non-aliasing.
\newblock In {\em PLDI 2003}, 2003.

\bibitem{BarwiseFeferman85ModelTheoreticLogics}
J.~Barwise and S.~Feferman, editors.
\newblock {\em Model-Theoretic Logics}.
\newblock Springer, 1985.

\bibitem{BenediktETAL99LogicForLinked}
M.~Benedikt, T.~Reps, and M.~Sagiv.
\newblock A decidable logic for linked data structures.
\newblock In {\em Proc. 8th ESOP}, 1999.

\bibitem{BerdineETAL04DecidableFragmentSeparationLogic}
J.~Berdine, C.~Calcagno, and P.~O'Hearn.
\newblock A decidable fragment of separation logic.
\newblock In {\em FSTTCS}, 2004.

\bibitem{BirkedalETAL04LocalReasoningCopyingCollector}
L.~Birkedal, N.~Torp-Smith, and J.~C. Reynolds.
\newblock Local reasoning about a copying garbage collector.
\newblock In {\em 31st ACM POPL}, pages 220--231. ACM Press, 2004.

\bibitem{Blanchet98EscapeAnalysis}
B.~Blanchet.
\newblock Escape analysis: correctness proof, implementation and experimental
  results.
\newblock In {\em Proceedings of the 25th ACM SIGPLAN-SIGACT symposium on
  Principles of programming languages}, pages 25--37. ACM Press, 1998.

\bibitem{Blanchet03EscapeAnalysis}
B.~Blanchet.
\newblock Escape {A}nalysis for {J}ava({TM}). {T}heory and {P}ractice.
\newblock {\em ACM Transactions on Programming Languages and Systems}, 2003.
\newblock To appear.

\bibitem{BoergerStaerk03AbstractStateMachines}
E.~B\"orger and R.~St\"ark.
\newblock {\em Abstract State Machines}.
\newblock Springer-Verlag, 2003.

\bibitem{Burstall72SomeTechniquesProvingCorrectnessPrograms}
R.~Burstall.
\newblock Some techniques for proving correctness of programs which alter data
  structures.
\newblock {\em Machine Intelligence}, 7, 1972.

\bibitem{Caires04BehavioralSpatialObservationsLogicPiCalculus}
L.~Caires.
\newblock Behavioral and spatial observations in a logic for the pi-calculus.
\newblock In {\em FoSSaCS}, 2004.

\bibitem{CairesCardelli03SpatialLogicConcurrencyPartI}
L.~Caires and L.~Cardelli.
\newblock A spatial logic for concurrency (part i).
\newblock {\em Information and Computation}, 186(2):194--235, 2003.

\bibitem{CalcagnoETAL02DecidingValiditySpatialLogicTrees}
C.~Calcagno, L.~Cardelli, and A.~D. Gordon.
\newblock Deciding validity in a spatial logic for trees.
\newblock In {\em ACM TLDI'02}, 2002.

\bibitem{CalcagnoETAL01ComputabilityComplexityResultsSpatialAssertion}
C.~Calcagno, H.~Yang, and P.~O'Hearn.
\newblock Computability and complexity results for a spatial assertion language
  for data structures.
\newblock In {\em FSTTCS}, 2001.

\bibitem{Ghelli02SpatialLogicMFPS}
L.~Cardelli, P.~Gardner, and G.~Ghelli.
\newblock A spatial logic for quering graphs.
\newblock Presentation at the 18th Workshop on Mathematical Foundations of
  Programming Semantics, March 2002.

\bibitem{LucaGhelli02SpatialLogicQueryingGraphs}
L.~Cardelli, P.~Gardner, and G.~Ghelli.
\newblock A spatial logic for querying graphs.
\newblock In {\em Proc. 29th ICALP}, volume 2380 of {\em LNCS}, 2002.

\bibitem{CardelliGhelli01QueryLanguageAmbientLogic}
L.~Cardelli and G.~Ghelli.
\newblock A query language based on the ambient logic.
\newblock In {\em Proc. 10th ESOP}, volume 2028 of {\em LNCS}, 2001.

\bibitem{CardelliGordon00AnytimeAnywhere}
L.~Cardelli and A.~D. Gordon.
\newblock Anytime, anywhere. modal logics for mobile ambients.
\newblock In {\em 27th ACM POPL}, 2000.

\bibitem{ComonETAL97Tata}
H.~Comon, M.~Dauchet, R.~Gilleron, F.~Jacquemard, D.~Lugiez, S.~Tison, and
  M.~Tommasi.
\newblock Tree automata techniques and applications.
\newblock Available on: {\tt http://www.grappa.univ-lille3.fr/tata}, 1997.
\newblock release 1999.

\bibitem{CousotCousot79SystematicDesignProgramAnalysis}
P.~Cousot and R.~Cousot.
\newblock Systematic design of program analysis frameworks.
\newblock In {\em Proc. 6th POPL}, pages 269--282, San Antonio, Texas, 1979.
  ACM Press, New York, NY.

\bibitem{DawarETAL04ExpressivenessComplexityGraphLogic}
A.~Dawar, P.~Gardner, and G.~Ghelli.
\newblock Expressiveness and complexity of graph logic.
\newblock Technical Report~3, Imperial College, Department of Computing
  Technical Report, 2004.

\bibitem{DeLineFahndrich01EnforcingHighLevelProtocols}
R.~DeLine and M.~F\"ahndrich.
\newblock Enforcing high-level protocols in low-level software.
\newblock In {\em Proc. ACM PLDI}, 2001.

\bibitem{FahndrichDeLine02AdoptionFocus}
M.~Fahndrich and R.~DeLine.
\newblock Adoption and focus: Practical linear types for imperative
  programming.
\newblock In {\em Proc. ACM PLDI}, 2002.

\bibitem{GraedelETAL97TwoVariableLogicCountingDecidable}
E.~Gr{\"a}del, M.~Otto, and E.~Rosen.
\newblock Two-variable logic with counting is decidable.
\newblock In {\em Proceedings of 12th IEEE Symposium on Logic in Computer
  Science LICS `97, Warschau}, 1997.

\bibitem{Hodges93ModelTheory}
W.~Hodges.
\newblock {\em Model Theory}, volume~42 of {\em Encyclopedia of Mathematics and
  its Applications}.
\newblock Cambridge University Press, 1993.

\bibitem{IshtiaqOHearn01BIAssertionLanguage}
S.~Ishtiaq and P.~W. O'Hearn.
\newblock {BI} as an assertion language for mutable data structures.
\newblock In {\em Proc. 28th ACM POPL}, 2001.

\bibitem{KodumalAiken04SetConstraintCFLReachabilityConnection}
J.~Kodumal and A.~Aiken.
\newblock The set constraint/{CFL} reachability connection in practice.
\newblock In {\em Proc. ACM PLDI}, June 2004.

\bibitem{KuncakETAL02RoleAnalysis}
V.~Kuncak, P.~Lam, and M.~Rinard.
\newblock Role analysis.
\newblock In {\em Proc. 29th POPL}, 2002.

\bibitem{KuncakRinard03OnRoleLogic}
V.~Kuncak and M.~Rinard.
\newblock On role logic.
\newblock Technical Report 925, MIT CSAIL, 2003.

\bibitem{KuncakRinard04FirstOrderTheorySetsCardinalityConstraintsDecidable}
V.~Kuncak and M.~Rinard.
\newblock The first-order theory of sets with cardinality constraints is
  decidable.
\newblock Technical Report 958, MIT CSAIL, July 2004.

\bibitem{KuncakRinard04GeneralizedRecordsRoleLogic}
V.~Kuncak and M.~Rinard.
\newblock Generalized records and spatial conjunction in role logic.
\newblock In {\em 11th Annual International Static Analysis Symposium
  (SAS'04)}, Verona, Italy, August 26--28 2004.

\bibitem{KuncakRinard04OnRecordsSpatialConjunctionRoleLogic}
V.~Kuncak and M.~Rinard.
\newblock On generalized records and spatial conjunction in role logic.
\newblock Technical Report 942, MIT CSAIL, April 2004.

\bibitem{LamETAL04OurExperienceModularPluggableAnalyses}
P.~Lam, V.~Kuncak, and M.~Rinard.
\newblock On our experience with modular pluggable analyses.
\newblock Technical Report 965, MIT CSAIL, September 2004.

\bibitem{LamETAL04HobProjectWebPage}
P.~Lam, V.~Kuncak, K.~Zee, and M.~Rinard.
\newblock The {H}ob project web page.
\newblock http://catfish.csail.mit.edu/$\sim$plam/hob/, 2004.

\bibitem{LevAmiETAL00PuttingStaticAnalysisWorkVerification}
T.~Lev-Ami, T.~Reps, M.~Sagiv, and R.~Wilhelm.
\newblock Putting static analysis to work for verification: A case study.
\newblock In {\em International Symposium on Software Testing and Analysis},
  2000.

\bibitem{LozesCaires04EliminationQuantifiersUndecidabilitySpatialLogicsConcurr%
ency}
{\'E}.~Lozes and L.~Caires.
\newblock Elimination of quantifiers and undecidability in spatial logics for
  concurrency.
\newblock In {\em CONCUR}, 2004.

\bibitem{ManevichETAL02CompactlyRepresentingFirstOrderStructures}
R.~Manevich, G.~Ramalingam, J.~Field, D.~Goyal, and M.~Sagiv.
\newblock Compactly representing first-order structures for static analysis.
\newblock In {\em Proc. 9th International Static Analysis Symposium}, pages
  196--212, 2002.

\bibitem{ManevichETAL04PartiallyDisjunctiveHeapAbstraction}
R.~Manevich, M.~Sagiv, G.~Ramalingam, and J.~Field.
\newblock Partially disjunctive heap abstraction.
\newblock In R.~Giacobazzi, editor, {\em Proceedings of the 11th International
  Symposium, SAS 2004}, volume 3148 of {\em Lecture Notes in Computer Science},
  pages 265--279. Springer, aug 2004.
\newblock Available at http://www.cs.tau.ac.il/$\sim$rumster/sas04.pdf.

\bibitem{MelskiReps00InterconvertibilitySetConstraintsCFLReachability}
D.~Melski and T.~Reps.
\newblock Interconvertibility of a class of set constraints and context-free
  language reachability.
\newblock {\em TCS}, 248:29--98, November 2000.

\bibitem{Mendelson97IntroductionMathematicalLogic}
E.~Mendelson.
\newblock {\em Introduction to Mathematical Logic}.
\newblock Chapman \& Hall, London, 4th edition, 1997.

\bibitem{Moeller01PALE}
A.~M{\o}ller and M.~I. Schwartzbach.
\newblock The {P}ointer {A}ssertion {L}ogic {E}ngine.
\newblock In {\em Proc. ACM PLDI}, 2001.

\bibitem{NielsonETAL01KleenesLogicEquality}
F.~Nielson, H.~R. Nielson, and M.~Sagiv.
\newblock Kleene's logic with equality.
\newblock {\em Inf. Process. Lett.}, 80(3):131--137, 2001.

\bibitem{OHearnPym99LogicBunchedImplications}
P.~O'Hearn and D.~Pym.
\newblock The logic of bunched implications.
\newblock {\em Bulleting of Symbolic Logic}, 5(2):215--244, 1999.

\bibitem{OHearnETAL01LocalReasoning}
P.~O'Hearn, J.~Reynolds, and H.~Yang.
\newblock Local reasoning about programs that alter data structures.
\newblock In {\em Proc. CSL, Paris 2001}, volume 2142 of {\em LNCS}, 2001.

\bibitem{Reynolds00IntuitionisticReasoningMutable}
J.~C. Reynolds.
\newblock Intuitionistic reasoning about shared mutable data structure.
\newblock In {\em Proceedings of the Symposium in Celebration of the Work of
  C.A.R. Hoare}, 2000.

\bibitem{Reynolds02SeparationLogic}
J.~C. Reynolds.
\newblock Separation logic: a logic for shared mutable data structures.
\newblock In {\em 17th LICS}, pages 55--74, 2002.

\bibitem{SagivETAL99Parametric}
M.~Sagiv, T.~Reps, and R.~Wilhelm.
\newblock Parametric shape analysis via 3-valued logic.
\newblock In {\em Proc. 26th ACM POPL}, 1999.

\bibitem{SagivETAL02Parametric}
M.~Sagiv, T.~Reps, and R.~Wilhelm.
\newblock Parametric shape analysis via 3-valued logic.
\newblock {\em ACM TOPLAS}, 24(3):217--298, 2002.

\bibitem{Skolem19Untersuchungen}
T.~Skolem.
\newblock Untersuchungen \"uber die {A}xiome des {K}lassenkalk\"uls and \"uber
  ``{P}roduktations- und {S}ummationsprobleme'', welche gewisse {K}lassen von
  {A}ussagen betreffen.
\newblock Skrifter utgit av {V}idnskapsselskapet i {K}ristiania, I. klasse, no.
  3, {O}slo, 1919.

\bibitem{SmithETAL00AliasTypes}
F.~Smith, D.~Walker, and G.~Morrisett.
\newblock Alias types.
\newblock In {\em Proc. 9th ESOP}, Berlin, Germany, Mar. 2000.

\bibitem{WalkerMorrisett00AliasTypesRecursive}
D.~Walker and G.~Morrisett.
\newblock Alias types for recursive data structures.
\newblock In {\em Workshop on Types in Compilation}, 2000.

\bibitem{WilhelmETAL00ShapeAnalysis}
R.~Wilhelm, M.~Sagiv, and T.~Reps.
\newblock Shape analysis.
\newblock In {\em Proc. 9th International Conference on Compiler Construction},
  Berlin, Germany, 2000. Springer-Verlag.

\bibitem{Yahav01VerifyingSafetyPropertiesConcurrentJava}
E.~Yahav.
\newblock Verifying safety properties of concurrent {J}ava programs using
  3-valued logic.
\newblock In {\em Proceedings of the 28th ACM SIGPLAN-SIGACT symposium on
  Principles of programming languages}, pages 27--40. ACM Press, 2001.

\bibitem{YahavETAL03VerifyingTemporalHeapPropertiesSpecifiedEvolutionLogic}
E.~Yahav, T.~Reps, M.~Sagiv, and R.~Wilhelm.
\newblock Verifying temporal heap properties specified via evolution logic.
\newblock In {\em Proc. 12th ESOP}, 2003.

\end{thebibliography}

\end{document}